\documentclass[aip, apl, reprint]{revtex4-1} 
\usepackage{lipsum}
\usepackage[version=3]{mhchem}
\usepackage{miller}
\usepackage{graphicx}
\usepackage{amsmath}
\usepackage{color}

\begin{document}
\title{Two-dimensional superconductivity at the \hkl(111)\ce{LaAlO3}/\ce{SrTiO3}interface} 

\author{A.M.R.V.L.~Monteiro}
\email{A.M.Monteiro@tudelft.nl}
\author{D.J.~Groenendijk}
\author{I.~Groen}
\author{J.~de Bruijckere}
\author{R.~Gaudenzi}
\author{H.S.J.~van der Zant}
\author{A.D.~Caviglia}
\email{A.Caviglia@tudelft.nl}
\affiliation{Kavli Institute of Nanoscience, Delft University of Technology,\\ P.O. Box 5046, 2600 GA Delft, Netherlands.}

\begin{abstract}
We report on the discovery and transport study of the superconducting ground state present at the \hkl(111)\ce{LaAlO3}/\ce{SrTiO3} interface. The superconducting transition is consistent with a Berezinskii-Kosterlitz-Thouless transition and its 2D nature is further corroborated by the anisotropy of the critical magnetic field, as calculated by Tinkham. The estimated superconducting layer thickness and coherence length are $10\,\mathrm{nm}$ and $60\,\mathrm{nm}$, respectively. The results of this work provide new insight to clarify the microscopic details of superconductivity in LAO/STO interfaces, in particular in what concerns the link with orbital symmetry.
\end{abstract}
\pacs{}

\maketitle 

Transition metal oxide interfaces host a rich spectrum of functional properties which are not present in their parent bulk constituents\,\cite{hwang2012emergent}. Following the groundbreaking discovery of a high-mobility two-dimensional electron system (2DES) at the interface between the two wide band-gap insulators \ce{LaAlO3} (LAO) and \ce{SrTiO3} (STO)\,\cite{ohtomo2004high}, a growing body of research efforts have brought to light many of its interesting properties. The system features a gate tunable metal-to-insulator transition\cite{thiel2006tunable,cen2008nanoscale}, strong Rashba spin-orbit coupling\,\cite{caviglia2010tunable} and superconductivity\cite{reyren2007superconducting}, possibly in coexistence with magnetism\cite{bert2011direct,li2011coexistence}.
To date, the vast majority of research efforts has been directed towards the investigation of the \hkl(001)-oriented LAO/STO interface. However, it is well recognized that the direction of confinement plays a pivotal role in determining hierarchy of orbital symmetries and, consequently, in properties such as the spatial extension of the 2DES and the Rashba spin-orbit fields\,\cite{herranz2015engineering}. 
Recent work suggests that \hkl(111)-oriented \ce{ABO3} perovskites are potentially suitable for the realization of topologically non-trivial phases\,\cite{xiao2011interface}, since along this direction a bilayer of \ce{B}-site ions forms a honeycomb lattice. The 2DES at the \hkl(111)LAO/STO interface\,\cite{herranz2012high} is an interesting subject of investigation, combining a polar discontinuity at the interface with such a hexagonal lattice. Signatures of the 6-fold symmetry related to the \hkl(111)STO orientation have recently been observed by angle-resolved photoemission spectroscopy\,\cite{rodel2014orientational,walker2014control} (ARPES) and magnetoresistance\cite{miao2016anisotropic,rout2017six} measurements, making the system potentially suitable for exotic time-reversal symmetry breaking superconductivity\cite{scheurer2017selection}. Moreover, ARPES measurements at the surface of \hkl(111)STO have confirmed a distinct orbital ordering of the $t_{2g}$ manifold\cite{rodel2014orientational}, where all the bands are degenerate at the $\Gamma$-point. This implies the absence of a Lifshitz point, considered to be at the origin of many physical properties at the \hkl(001)-oriented interface. In particular, the `optimal doping' for superconductivity was found to occur concomitantly with the Lifshitz transition\,\cite{joshua2011universal}.
Therefore, within this view, it is timely to investigate whether a 2D superconducting ground state arises at the \hkl(111) orientation.

\begin{figure}
\includegraphics[width=\linewidth]{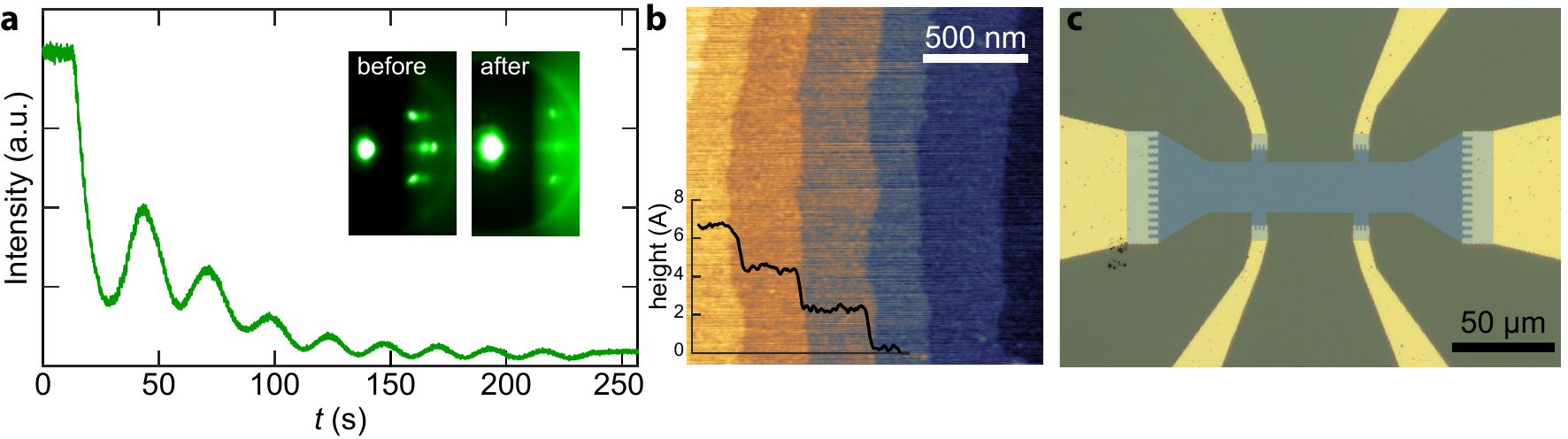}%
\caption{
(a) RHEED intensity oscillations of the specular spot during the epitaxial growth of a 9~u.c.~\ce{LaAlO3} film on a \hkl(111)\ce{SrTiO3} substrate. Inset: RHEED pattern before and after growth. 
(b) AFM topographic image of the surface after growth. Inset: height profile.
(c) Optical image of a Hall bar. The channel is false-coloured in blue.
\label{fig:fab}}%
\end{figure}

The \hkl(111)LAO/STO interface under investigation was prepared by pulsed laser deposition. An LAO film with a thickness of 9 unit cells (u.c.)~was epitaxially grown on a commercially available \hkl(111)STO substrate with Ti-rich surface. The film was deposited at $840^{\circ}\mathrm{C}$ in an oxygen pressure of $6\times 10^{-5}\,\mathrm{mbar}$. The laser pulses were supplied by a KrF excimer source $(\lambda=248\,\mathrm{nm})$ with an energy density of $1\,\mathrm{J/cm^{2}}$ and a frequency of $1\,\mathrm{Hz}$. The growth process was followed by an annealing step in order to refill oxygen vacancies. 
The chamber was filled with $300\,\mathrm{mbar}$ of oxygen and the sample temperature was kept at $600^{\circ}\mathrm{C}$ for 1 hour. The sample was then cooled down to room temperature at a rate of $10^{\circ}\,\mathrm{C/min}$ in the same oxygen atmosphere. The growth process was monitored \textit{in-situ} using reflection high-energy electron diffraction (RHEED), which indicates a layer-by-layer growth mode, as shown in Figure\,\ref{fig:fab}a. An atomic force microscope (AFM) topographic image of the surface after growth is shown in Figure\,\ref{fig:fab}b, where an atomically flat surface with step-and-terrace structure can be observed. The step height corresponds to the \hkl(111)STO interplanar distance ($\approx0.26\,\mathrm{\AA}$). Transport measurements were carried out in a Hall bar geometry, as shown in Figure\,\ref{fig:fab}c. The fabrication process relied on argon dry etching in order to define the channel and e-beam evaporation of metal contacts (for a detailed description, see Supporting Information). Hall bars were patterned along different in-plane orientations ($0^{\circ}$, $30^{\circ}$, $60^{\circ}$, and $90^{\circ}$) in order to investigate possible anisotropies in the transport properties.

%
\begin{figure}[ht!]
\includegraphics[trim={3.5cm 0 4cm 0},clip,width=\linewidth]{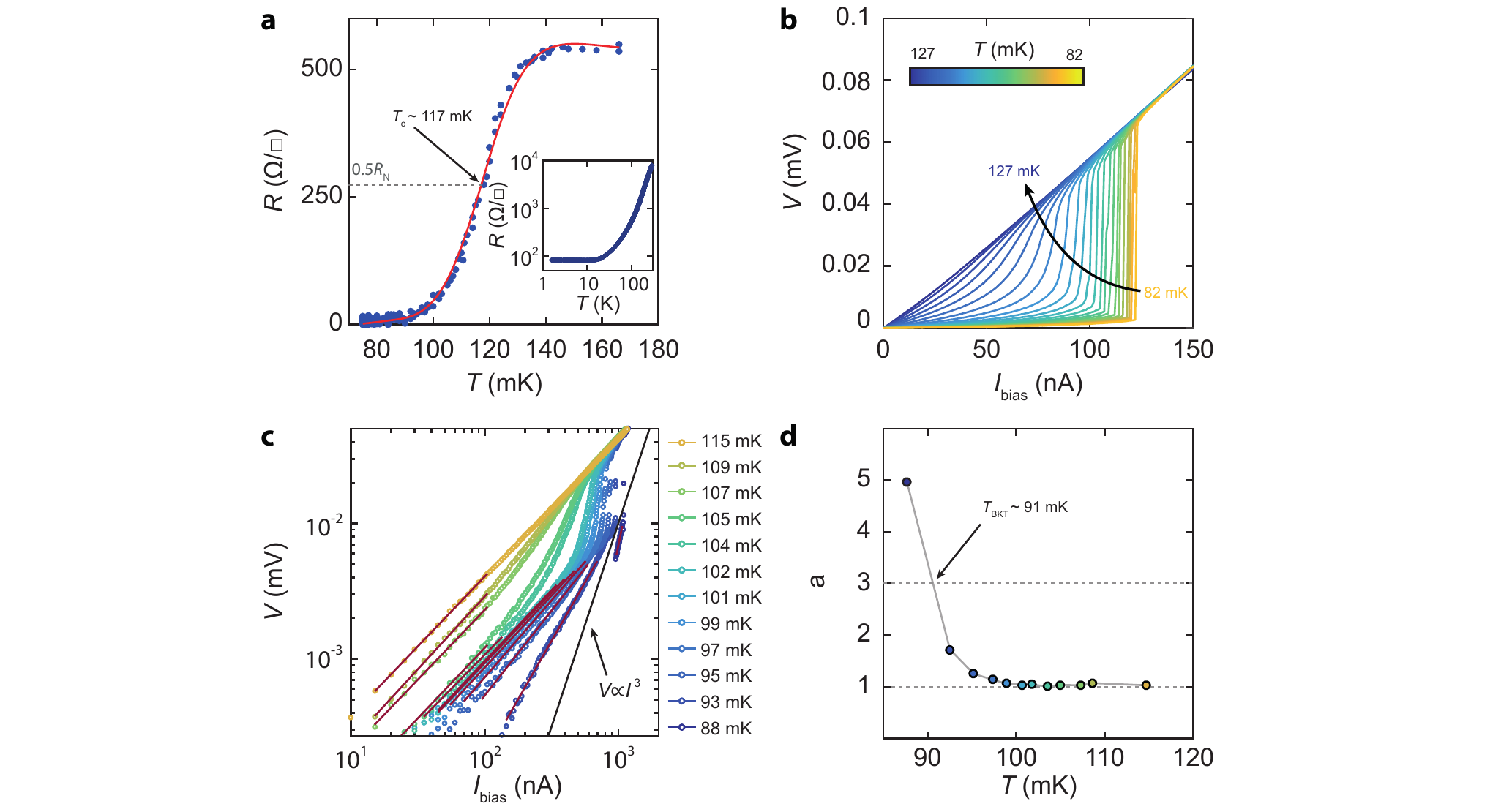}%
\caption{
(a) Sheet resistance ($R$) as a function of temperature ($T$) in the milliKelvin regime with an applied back gate voltage of $30\,\mathrm{V}$. Inset: $R(T)$ in the high-temperature range down to $1.5\,\mathrm{K}$.
(b) $V(I)$ characteristics measured at different temperatures. 
(c) Selected curves of panel (b) plotted in logarithmic scale. The red lines are fits of the data along the transition. The black line corresponds to $V\propto I^3$.
(d) Temperature dependence of the power-law exponent $a(T)$ as deduced from the fits shown in (c). 
\label{fig:bkt}}%
\end{figure}

The temperature dependence of the sheet resistance ($R$) is shown in Figure\,\ref{fig:bkt}a, evidencing a clear metallic behavior and absence of carrier localisation down to $1.5\,\mathrm{K}$. At this temperature the back gate voltage is swept to the maximum applied voltage ($90\,\mathrm{V}$) and back to $0\,\mathrm{V}$. At variance with previous reports, we observed no hysteretic or anisotropic transport behavior attributed to the presence of oxygen vacancies\,\cite{davis2016electrical}.  
All further measurements presented in this work are shown for one Hall bar recorded at a fixed back gate voltage of $30\,\mathrm{V}$. The detailed investigation of the evolution of the transport properties with electrostatic doping shall be discussed elsewhere. 

In the milliKelvin regime, a superconducting transition with a critical temperature $T_{\mathrm{c}}\approx 117\,\mathrm{mK}$ is observed (Figure\,\ref{fig:bkt}a). The value of $T_{\mathrm{c}}$ was defined as the temperature at which the resistance is 50\% of its normal state value ($R_{\mathrm{n}}$, measured at $T=180\,\mathrm{mK}$). The width of the transition, defined between 20\% and 80\% of $R_{\mathrm{n}}$, is $\Delta T_{\mathrm{c}}=17\,\mathrm{mK}$.

For a 2D system, it is well established that superconductivity should exhibit a Berezinskii-Kosterlitz-Thouless (BKT) transition, at a characteristic temperature $T_{\mathrm{BKT}}$. Below this temperature, vortex-antivortex pairs are formed. As the temperature increases and approaches $T_{\mathrm{BKT}}$, a thermodynamic instability occurs and the vortex-antivortex pairs spontaneously unbind into free vortices. The resulting  proliferation of free vortices destroys superconductivity, yielding a finite-resistance state. According to the BKT scenario, a strong non-Ohmic behavior in the $V(I)$ characteristics emerges near $T_{\mathrm{BKT}}$, following a power law behavior $V \propto I^{a(T)}$ with $a=3$ at $T_{\mathrm{BKT}}$.

In order to investigate the 2D character of superconductivity in the system, we measured the $V(I)$ characteristics of a 9~u.c. \hkl(111)LAO/STO interface as a function of temperature. The characteristics were recorded from $82\,\mathrm{mK}$, where the samples are completely superconducting, up to the temperature at which  the sample fully recovers to the normal state. As shown in Figure\,\ref{fig:bkt}b, there is a clear superconducting current plateau for the $V(I)$ curve at $82\,\mathrm{mK}$. As the temperature is increased, the supercurrent plateau becomes progressively  shorter, until it vanishes at approximately $127\,\mathrm{mK}$. At this temperature,  the $V(I)$ curve becomes completely linear. 
Concomitantly with the disappearance of the superconducting plateau, power-law type $V(I)$ curves emerge, indicating a BKT transition. In order to confirm this scenario, we plot the $V(I)$ characteristics  on a logarithmic scale, as shown in Figure\,\ref{fig:bkt}c. At sufficiently low current, the $V(I)$ characteristics display Ohmic behaviour in the entire temperature range due to well-known finite size effects\,\cite{kosterlitz1972long,kosterlitz1973ordering}. At higher current values, the $V(I)$ curves show a clear $V\propto I^{a(T)}$ power-law dependence, as indicated by the red lines. The black line corresponds to $V\propto I^3$. The exponents $a(T)$ are obtained by fitting all the characteristics and are plotted as a function of temperature in Figure\,\ref{fig:bkt}d, revealing that $T_{\mathrm{BKT}} \approx 91\,\mathrm{mK}$. At $T>T_{\mathrm{BKT}}$, $V \propto I$ at low currents, transitioning to a strongly non-linear behaviour at higher currents and showing the characteristic rounding. 
In contrast, at $T<T_{\mathrm{BKT}}$ the power-law behaviour terminates abruptly with a voltage jump at a well defined current. It should be noted that the evolution of $a(T)$ does not display the characteristic discontinuous jump from $a(T^{+}_{\mathrm{BKT}})=1$ to $a(T^{-}_{\mathrm{BKT}})=3$, but rather transitions smoothly from 1 to 3 over a range of several milliKelvin. This behaviour, also observed in \hkl(001)- and \hkl(110)-oriented interfaces\,\cite{reyren2007superconducting,han2014two}, stems from inhomogeneties in the local superconducting properties of the system (such as inhomogeneous superfluid density\,\cite{bert2012gate} or structural twin domains of the STO substrate\,\cite{noad2016variation}) which smear the universal jump\,\cite{baity2016effective}.

\begin{figure}[ht]
\includegraphics[trim={3.5cm 0 4cm 0},clip,width=\linewidth]{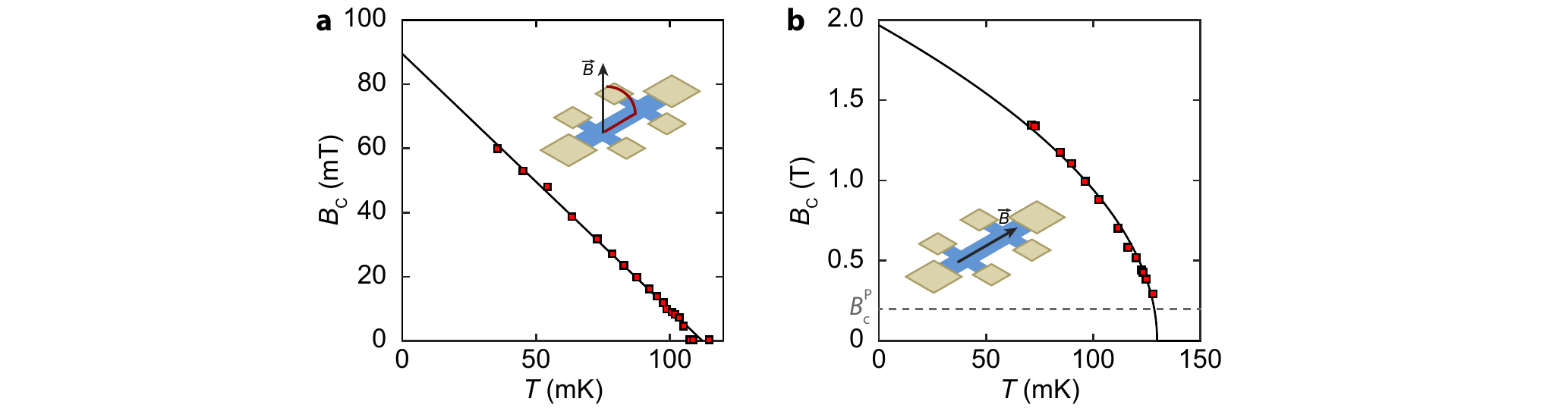}
\caption{
(a) Out-of-plane and (b) in-plane critical magnetic field as a function of temperature. Dashed line: Pauli-limiting field.
\label{fig:Bc}}%
\end{figure}

For a quantitative estimation of both the superconducting coherence length ($\xi$) and the layer thickness $d$, we carried out an analysis based on the Landau-Ginzburg formalism. To this purpose, a quantitative criterion was chosen in order to determine the out-of-plane ($B^{\perp}_{\mathrm{c}}$) and in-plane ($B^{\parallel}_{\mathrm{c}}$)  critical magnetic fields. 
At each temperature, $V(I)$ characteristics are recorded for increasing values of applied magnetic field. 
As shown in the Supporting Information, for small applied magnetic fields, the values of $dV/dI_{I=0\,\mathrm{nA}}$ are zero at low currents, corresponding to the superconducting state. As the current rises, $dV/dI$ increases until a saturating value, $dV/dI_{I=200\,\mathrm{nA}}$, which corresponds to the normal state resistance. For larger applied magnetic fields, $dV/dI_{I=0\,\mathrm{nA}}$ is non-zero, and its value increases with the magnitude of the applied magnetic field.

We define the critical magnetic field as the value at which $dV/dI_{I=0\,\mathrm{nA}}$ reaches $50\%$ of the normal state resistance, i.e., $(dV/dI)_{I=0\,\mathrm{nA}}/(dV/dI)_{I=200\,\mathrm{nA}}=€‰0.5$.

We track the temperature evolution of the critical magnetic field for the out-of-plane and in-plane orientations, which are shown in Figure\,\ref{fig:Bc}a and b, respectively. The black lines represent a fit to the expected dependence for a 2D superconductor, i.e.,

\begin{equation}
B^{\perp}_{\mathrm{c}} = \frac{\Phi_0}{2 \pi \xi^2}(1-T/T_{\mathrm{c}})
\end{equation}

and

\begin{equation}
B^{\parallel}_{\mathrm{c}} = \frac{\Phi_0 \sqrt{12}}{2 \pi \xi d}(1-T/T_{\mathrm{c}})^{1/2}.
\end{equation}

From the extrapolation of the critical magnetic fields at $T=0\,\mathrm{K}$, we extracted the in-plane coherence length $\xi = \sqrt{\frac{\Phi0}{2 \pi B^{\perp}_{\mathrm{c,0K}}}} \approx 60\,\mathrm{nm}$ and the thickness of the superconducting layer $d = \frac{\Phi_0 \sqrt{3}}{\pi \xi B^{\parallel}_{\mathrm{c,0K}}} \approx 10\,\mathrm{nm}$. The fact that the superconducting coherence length is larger than the estimated thickness is consistent with the 2D character of superconductivity.

In fact, $B_{\mathrm{c}}^{\parallel}$ can seemingly go far beyond the Pauli paramagnetic limit, which gives a higher bound for the upper critical magnetic field resulting from field-induced pair-breaking\,\cite{chandrasekhar1962note,clogston1962upper}. For weak coupling Bardeen-Cooper-Schrieffer (BCS) superconductors, this value is given by 

\begin{equation}
B^{\mathrm{P}}_{\mathrm{c}} \approx 1.76 k_{\mathrm{B}} T_{\mathrm{c}}/\sqrt{2} \mu_{\mathrm{B}},
\end{equation}

where $k_{\mathrm{B}}$ is the Boltzmann's constant and $\mu_{\mathrm{B}}$ is the Bohr magneton. 

Violation of the paramagnetic limit has been observed in \hkl(001)- and \hkl(110)-oriented LAO/STO interfaces\cite{han2014two,reyren2009anisotropy,shalom2010tuning}, as well as in other STO-based superconductors\,\cite{kim2012intrinsic}. In these systems, the paramagnetic limit is exceeded by a factor of approximately $3$-$5$.
In our case, we find $B^{\mathrm{P}}_{\mathrm{c}} \approx 200\,\mathrm{mT}$, which results in a violation of the Pauli paramagnetic limit by a factor of 10, since $B^{\parallel}_{\mathrm{c,0K}} \approx 2000\,\mathrm{mT}$. As a matter of fact, the violation is already present at temperatures very close to $T_{\mathrm{c}}$, as shown by the dashed line in Figure\,\ref{fig:Bc}b. 
The enhancement of $B^{\parallel}_{\mathrm{c}}$ well beyond the BCS prediction has been reported in superconductors which display strong spin-orbit effects\,\cite{lu2013superconductivity,khim2011pauli,gardner2011enhancement}. These are expected to cause randomization of electron spins, and thus result in suppression of the effect of spin paramagnetism\,\cite{maki1966effect}.
Indeed, we have confirmed the presence of strong spin-orbit fluctuations in the system by magnetotransport measurements (see Supporting Information), suggesting that spin-orbit coupling plays an important role in the violation of the Pauli paramagnetic limit.

\begin{figure}
\includegraphics[trim={3.5cm 0 4cm 0},clip,width=\linewidth]{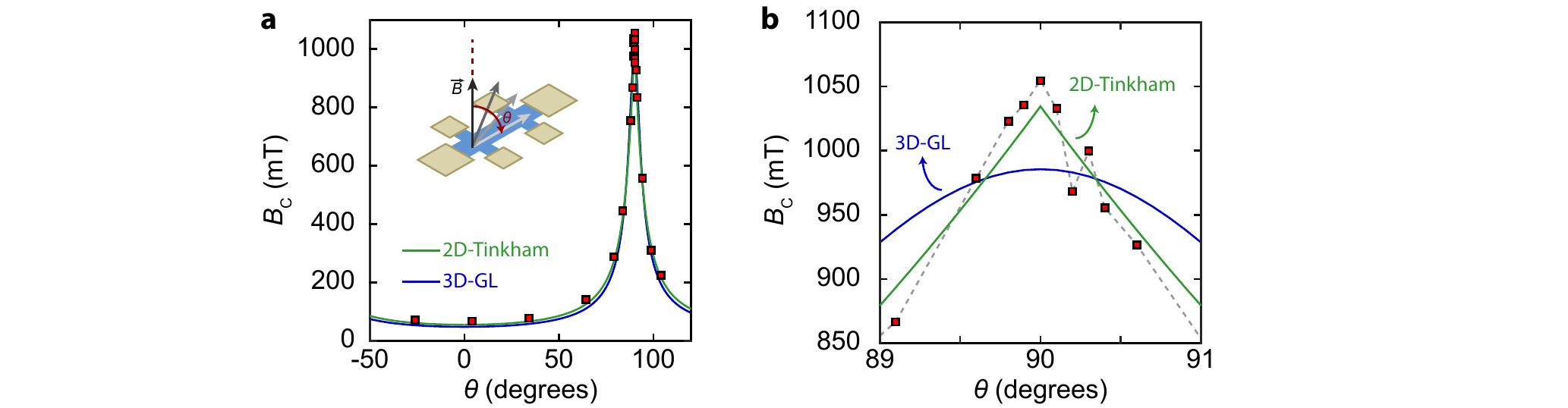}
\caption{
(a) Angular dependence of the critical magnetic field $B_{\mathrm{c}}$, where $\theta$ is the angle between the magnetic field and the surface normal. Green line: fit to the 2D Tinkham formula. Blue: fit to the 3D anisotropic Ginzburg-Landau model. (b) Magnified view of the region around $\theta = 90^{\circ}$.
\label{fig:angular}}%
\end{figure}

To further investigate the dimensionality of the superconducting layer, we have studied the angular dependence of the critical magnetic field at $T=82\,\mathrm{mK}$. Figure\,\ref{fig:angular}a shows the critical magnetic field as a function of the angle $\theta$, defined between the magnetic field vector and the normal to the surface.
The data was fitted with the 2D Tinkham formula (green curve) and the 3D anisotropic Ginzburg-Landau model (blue curve), given by

\begin{equation}
\frac{B^{\theta}_{\mathrm{c}}\left|\cos(\theta)\right|}{B^{\perp}_{\mathrm{c}}} + \left(\frac{B^{\theta}_{\mathrm{c}}\sin(\theta)}{B^{\parallel}_{\mathrm{c}}}\right)^2 =1
\end{equation}

and

\begin{equation}
\left(\frac{B^{\theta}_{\mathrm{c}}\cos(\theta)}{B^{\perp}_{\mathrm{c}}}\right)^2 + \left(\frac{B^{\theta}_{\mathrm{c}}\sin(\theta)}{B^{\parallel}_{\mathrm{c}}}\right)^2 =1,
\end{equation}

respectively.

For the overall range, the data seems to be well described by both models. However, a closer look at the region around $\theta=90^{\circ}$ in Figure\,\ref{fig:angular}b reveals a clear difference between the two models: the 3D model yields a rounded maximum when the magnetic field vector is completely in plane, while the observed cusp-shaped peak can only be well captured by the 2D model.

In summary, by means of systematic (magneto)transport measurements we have demonstrated that the electrons hosted at the \hkl(111)LAO/STO interface condense into a superconducting ground state with $T_{\mathrm{c}} \approx 117\,\mathrm{mK}$. The estimated thickness of the 2D superconducting layer is approximatelly $10\,\mathrm{nm}$, very similar to the one usually reported for \hkl(001)-oriented interfaces. The $V(I)$ characteristics are consistent with a BKT transition, and the two-dimensional character of the superconducting layer was further corroborated by the angular dependence of the critical magnetic field. The Pauli paramagnetic limit is exceeded by a factor of 10, indicating strong spin-orbit coupling in the system. 
In view of the differences between the symmetries, electronic structure, and orbital ordering of the confined states at the \hkl(001)- and \hkl(111)-oriented LAO/STO interfaces, further investigation of the latter can extend the current understanding of the link between orbital symmetry and superconductivity at LAO/STO interfaces.

\begin{acknowledgments}
We thank T.~Baturina for fruitful discussions on the BKT analysis and T.~Kool for technical support.
This work was supported by The Netherlands Organisation for Scientific Research (NWO/OCW) as part of the Frontiers of Nanoscience program (NanoFront), by the Dutch Foundation for Fundamental Research on Matter (FOM). The research leading to these results has received funding from the European Research Council under the European Union's H2020 programme/ ERC GrantAgreement n.~[677458].
\end{acknowledgments}

\bibliography{references}
\end{document}